\documentclass[prd,12pt,nofootinbib]{revtex4-1}

\usepackage{amsmath}
\usepackage{subfigure}
\usepackage{amsthm}
\usepackage{commath}
\usepackage{graphicx}
\usepackage{amssymb}
\usepackage{bm}
\usepackage{slashed}
\usepackage{tikz}
\usepackage{braket}
\usepackage[labelsep=space]{caption}
\def\fcncquark{\mbox{$b\to s\ell^+\ell^-$}}
\def\processKstar{\mbox{ $B \to K^{*} \mu^+ \mu^-$ }}
\def\processK{\mbox{ $B \to K \mu^+ \mu^-$ }}
\def\processphi{\mbox{ $B_s \to \phi \mu^+ \mu^-$ }}
\def\be{\begin{equation}}
\def\ee{\end{equation}}	
\def\bea{\begin{eqnarray}}
\def\eea{\end{eqnarray}}

\def\beq{\begin{equation}}
\def\eeq{\end{equation}}

\def\rar{\rightarrow}

\def\ba{\begin{array}}
	\def\ea{\end{array}}

\def\tepm{$B \rar K^* \mu^+ \mu^-$} 
 
\def\dis{\displaystyle}

\usepackage{hyperref}
\usepackage{epstopdf}

\begin{document}
\title{Re-examination of the rare decay $B_s\to\phi\mu^+\mu^-$ using holographic light-front QCD}

\author{Mohammad Ahmady}

\email{mahmady@mta.ca}
\affiliation{\small Department of Physics, Mount Allison University, \mbox{Sackville, New Brunswick, Canada, E4L 1E6}}
\author{Spencer Keller}
\email{stkeller@mta.ca}
\affiliation{\small Department of Physics, Mount Allison University, \mbox{Sackville, New Brunswick, Canada, E4L 1E6}}
\author{Michael Thibodeau}
\email{methibodeau@mta.ca}
\affiliation{\small Department of Physics, Mount Allison University, \mbox{Sackville, New Brunswick, Canada, E4L 1E6}}
\author{Ruben Sandapen}
\email{ruben.sandapen@acadiau.ca}
\affiliation{\small Department of Physics, Acadia University,
	 \mbox{Wolfville, Nova-Scotia, Canada, B4P 2R6}}
\affiliation{\small Department of Physics, Mount Allison University, \mbox{Sackville, New Brunswick, Canada, E4L 1E6}}

\begin{abstract}
 We calculate the Standard Model (SM) predictions for the differential branching ratio of the rare \processphi decays using $B_s \to \phi$ transition form factors (TFFs) obtained using holographic light-front QCD (hQCD) instead of the traditional QCD sum rules (QCDSR) .  Our predictions for the differential branching ratio is in better agreement with the LHCb data.  Also, we find that the hQCD prediction for $R_{K^*\phi}$, the ratio of the branching fraction of \processKstar to that of \processphi , is in excellent agreement with both the LHCb and CDF results in low $q^2$ range.    
 
\end{abstract}


\maketitle

\section{Introduction}
\begin{figure}
 	\centering
 	\subfigure[\mbox{ }Penguin diagram]{\includegraphics[width=0.3\textwidth]{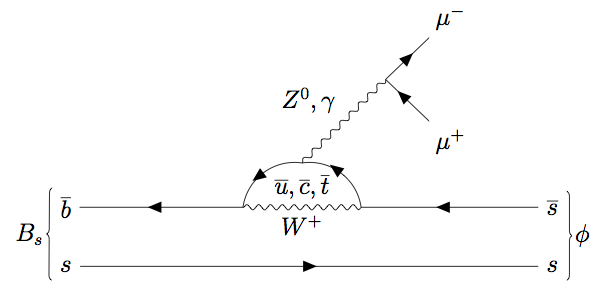}}
 	\subfigure[\mbox{ }Box diagram]{\includegraphics[width=0.3\textwidth]{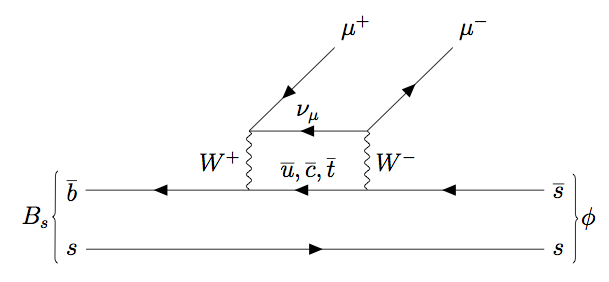}}
 	\caption{Feynman diagrams of the principal contributions to the $B_s\to \phi\mu^+\mu^-$ decay.}
 	\label{fig:feynmandiagrams}
 \end{figure}

 The flavor changing neutral current (FCNC) $b\to s$ transition, which is forbidden at tree-level, has been at the focus of extensive experimental and theoretical investigations.  This is due to the fact that, among other things, this rare transition is sensitive to New Physics (NP), i.e. physics beyond the Standard Model (SM).  In extensions of the SM, new heavy particles can appear in competing diagrams and affect both the branching fraction of the decay
 and the angular distributions of the final-state particles.  In particular, the semi-leptonic \fcncquark quark decay has received significant attention via measurements of inclusive $B\to X_s\ell^+\ell^-$ and exclusive \processK and \processKstar decays and their comparison against the SM predictions. Many observables for the dileptonic \tepm decay have already been measured and the precision of the experimental data is expected to improve significantly in the near future. The decay \processphi, which is closely related to the decay \processKstar, provides an alternate venue to examine the same underlying quark process, as shown in Fig.\ref{fig:feynmandiagrams}, in a different hadronic environment.   
 This decay channel was first observed and studied by the CDF collaboration \cite{Aaltonen:2011cn,Aaltonen:2011qs} and subsequently studied by the LHCb collaboration using data collected during 2011, corresponding to an integrated luminosity of $1.0$ fb$^{-1}$ \cite{Aaij:2013aln}.  Although $B_s$ meson production is suppressed with respect to the $B_d$ meson by the fragmentation fraction ratio $f_s/f_d\sim 1/4$, the narrow $\phi$ resonance allows a clean selection with low background levels. While the angular distributions were found to be in good agreement with SM expectations, the measured branching fraction deviates from the recently updated SM prediction by $3.1\,\sigma$ \cite{Altmannshofer:2014rta,Aaij:2015esa,Straub:2015ica,Geng:2003su}:
 \beq
 \label{data:phimumu}
 \frac{d~}{dq^2}{\rm BR}(B_s\to\phi\mu^+\mu^-){\Big|}_{q^2\in[1:6]\,{\rm GeV}^2}
 = \left\{ 
 \ba{lcl}
 \dis \left(2.58^{+0.33}_{-0.31}\pm 0.08\pm 0.19\right) \times 10^{-8}~{\rm GeV}^{-2} & \quad & ({\rm exp.}) \\[2ex]
 \dis \left(4.81 \pm 0.56\right) \times 10^{-8}~{\rm GeV}^{-2} & & ({\rm SM})\,,
 \ea
 \right.
 \eeq
   where $q^2 = m^2_{\mu\mu}$ is the invariant mass of the produced di-muons. A similar trend is also seen for the branching fractions of other $b\to s\mu^+ \mu^-$ processes, which tend to be lower than the SM predictions \cite{Aaij:2013iag,Aaij:2014pli,Aaij:2014kwa}. 
   
   One important aspect of SM theoretical predictions of the exclusive decays is the computation of the TFFs which parametrize the hadronic matrix elements of $B_{(s)}$ to light mesons through quark currents.  These nonperturbative TFFs are commonly evaluated using light-cone sum rules (LCSR)\cite{Ali:1993vd} with the input distribution amplitudes (DAs) obtained via QCDSR.  An alternative is to calculate these DAs using the light-front wavefunction of the light meson bound state.  In a recent work, we have shown that the light-front wavefunctions for $\rho$ and $\phi$ mesons resulting from hQCD leads to predictions for diffractive production cross sections of these vector mesons that are in good agreement with the experimental data\cite{Ahmady:2016ujw}.  This is our motivation to use the hQCD DAs to calculate the $B_s\to\phi$ TFFs and consequently, provide alternative predictions for the \processphi differential branching ratio.  The nonperturbative TFFs are the dominant source of the theoretical uncertainty in this decay mode and therefore it is important to improve our understanding of the corresponding error by examining viable models.

 \section{Holographic meson wavefunctions}
 \label{Holographic wfn}
 
 In recent years, new insights about hadronic light-front wavefunctions based on the anti-de Sitter/Conformal Field Theory (AdS/CFT) correspondence have been proposed by Brodsky and de T\'eramond \cite{deTeramond:2005su,Brodsky:2006uqa,deTeramond:2008ht}. In a semiclassical approximation of light-front QCD where quark masses and quantum loops are neglected, the meson wavefunction can be written as \cite{Brodsky:2014yha} 
 
 \begin{equation}
 	\Psi( x,\zeta, \phi)= e^{iL\phi} \mathcal{X}(x) \frac{\phi (\zeta)}{\sqrt{2 \pi \zeta}} 
 	\label{mesonwf}
 \end{equation}
 where the variable $\zeta=\sqrt{x(1-x)} r$ is the transverse separation between the quark and the antiquark at equal light-front time. The transverse wavefunction $\phi(\zeta)$ is a solution of the so-called holographic light-front Schr\"odinger equation:
 \begin{equation}
 	\left(-\frac{d^2}{d\zeta^2} - \frac{1-4L^2}{4\zeta^2} + U(\zeta) \right) \phi(\zeta) = M^2 \phi(\zeta) 
 	\label{holograhicSE}
 \end{equation}
 where $M$ is the mass of the meson and $U(\zeta)$ is the confining potential which at present cannot be computed from first-principles in QCD. On the other hand, making the substitutions $\zeta \to z$ where $z$ being the fifth dimension of AdS space, together with   $L^2 -(2-J)^2 \to (mR)^2$  where $R$ and $m$ are the radius of curvature and mass parameter of AdS space respectively, Eq. \eqref{holograhicSE} describes the propagation of spin-$J$ string modes in 5-D AdS space. In this case, the potential is given by
 \begin{equation}
 	U(z, J)= \frac{1}{2} \varphi^{\prime\prime}(z) + \frac{1}{4} \varphi^{\prime}(z)^2 + \left(\frac{2J-3}{4 z} \right)\varphi^{\prime} (z) 
 \end{equation}
 where $\varphi(z)$ is the dilaton field which breaks the conformal invariance of AdS space. A quadratic dilaton ($\varphi(z)=\kappa^2 z^2$) profile results in a harmonic oscillator potential in physical spacetime:
 \begin{equation}
 	U(\zeta,J)= \kappa^4 \zeta^2 + \kappa^2 (J-1) \;.
 	\label{harmonic-LF}
 \end{equation}
 The choice of a quadratic dilaton is dictated by the de Alfaro,
 Furbini and Furlan (dAFF)\cite{deAlfaro:1976vlx} mechanism for breaking conformal symmetry in the Hamiltonian while retaining the conformal invariance of the action\cite{Brodsky:2013npa}.  Solving the holographic Schr\"odinger equation with this harmonic potential given by Eq. \eqref{harmonic-LF} yields the meson mass spectrum,
 \begin{equation}
 	M^2= 4\kappa^2 \left(n+L +\frac{S}{2}\right)\;
 	\label{mass-Regge}
 \end{equation}
 with the corresponding normalized eigenfunctions
 \begin{equation}
 \phi_{n,L}(\zeta)= \kappa^{1+L} \sqrt{\frac{2 n !}{(n+L)!}} \zeta^{1/2+L} \exp{\left(\frac{-\kappa^2 \zeta^2}{2}\right)} L_n^L(x^2 \zeta^2) \;.
 \label{phi-zeta}
 \end{equation}
 To completely specify the holographic wavefunction given by Eq. \eqref{mesonwf}, the longitudinal wavefunction $\mathcal{X}(x)$ must be determined. For massless quarks, this is achieved by an exact mapping of the pion  electromagnetic form factors in AdS and in physical spacetime resulting in  \cite{Brodsky:2014yha}.
 \begin{equation}
 \mathcal{X}(x)=\sqrt{x(1-x)}
 \end{equation}
 
 Equation \eqref{mass-Regge} predicts that the mesons lie on linear Regge trajectories as  is experimentally observed and thus $\kappa$ can be chosen to fit the Regge slope. Ref. \cite{Brodsky:2014yha} reports $\kappa=0.54$ GeV for vector mesons. Eq. \eqref{mass-Regge} also predicts that the pion and kaon (with $n=0, L=0, S=0$) are massless.  For the ground state mesons with $n=0, L=0$,  Eq. \eqref{mesonwf} becomes
 \be 
 \Psi (x,\zeta)= \frac{\kappa}{\sqrt{\pi}} \sqrt {x(1-x) }  \exp{\left[-{\kappa^2 \zeta^2 \over 2} \right] } \;.
 \label{wavef}
 \ee
 To account for non-zero quark masses, we follow the prescription of Brodsky and de T\'eramond given in Ref. \cite{Brodsky:2008pg} which leads to an augmented form for the transverse part of the light-front wavefunction:
 \be  \Psi^\phi_{\lambda} (x,\zeta) = \mathcal{N}_{\lambda} \sqrt{x (1-x)}  \exp{ \left[ -{ \kappa^2 \zeta^2  \over 2} \right] }
 \exp{ \left[ -{m_s^2 \over 2 \kappa^2 x(1-x) } \right]}\,,
 \label{hwf}
 \ee
 where we have introduced a polarization-dependent normalization constant ${\mathcal N}_{\lambda}$ where $\lambda =L,\; T$.  Including the spin structure, the vector meson light-front wavefunctions can be written as \cite{Forshaw:2012im}
 \be
 \Psi^{\phi,L}_{h,\bar{h}}(x, r) =  \frac{1}{2} \delta_{h,-\bar{h}}  \bigg[ 1 + 
 { m_{s}^{2} -  \nabla_r^{2}  \over x(1-x)M^2_{V} } \bigg] \Psi^\phi_L(x, \zeta) \,,
 \label{mesonL}
 \ee
 and
 \be \Psi^{\phi, T}_{h,\bar{h}}(x, r) = \pm \bigg[  i e^{\pm i\theta_{r}}  ( x \delta_{h\pm,\bar{h}\mp} - (1-x)  \delta_{h\mp,\bar{h}\pm})  \partial_{r}+ m_{s}\delta_{h\pm,\bar{h}\pm} \bigg] {\Psi^\phi_T(x, \zeta) \over 2 x (1-x)}\,. 
 \label{mesonT}
 \ee
 
 We fix the normalization constant $\mathcal{N}_{\lambda}$ in Eq. \eqref{hwf} by requiring that
 \be
 \sum_{h,\bar{h}} \int {\mathrm d}^2 {\mathbf{r}} \, {\mathrm d} x |
 \Psi^{\phi, \lambda} _{h, {\bar h}}(x, r)|^{2} = 1 \,.
 \ee
 
Previously, we have used holographic light-front wavefunctions to calculate hadronic effects in rare B decays to $\rho$ and $\phi$\cite{Ahmady:2018fvo,Ahmady:2015fha,Ahmady:2014cpa,Ahmady:2014sva,Ahmady:2013cva,Ahmady:2012dy}.  
 
\section{decay constants and distribution amplitudes}
 Having specified the holographic wavefunction for $\phi$ meson, we are now able to predict their vector and tensor couplings, $f_\phi$ and $f_\phi^T$ respectively,  defined by \cite{Ball:1998sk}
 \begin{equation}
 	\langle 0|\bar s(0)  \gamma^\mu s(0)|\phi
 	(P,\lambda)\rangle =f_\phi M_\phi e_\lambda^{\mu}
 	\label{fv-def}
 \end{equation}
 and
 \begin{equation}
 	\langle 0|\bar s(0) [\gamma^\mu,\gamma^\nu] s(0)|\phi (P,\lambda)\rangle =2 f_\phi^{T}  (e^{\mu}_{\lambda} P^{\nu} - e^{\nu}_{\lambda} P^{\mu}) \;.
 	\label{fvT-def}
 \end{equation}
 respectively.  In Eqs. \eqref{fv-def} and \eqref{fvT-def}, the strange quark and antiquark fields evaluated at the same spacetime point, $P^\mu$ and $e^{\mu}_{\lambda}$ are the momentum and polarization vectors of the $\phi$ meson.  It follows that\cite{Ahmady:2012dy,Ahmady:2016ujw}
 \bea
 f_{\phi} &=&  {\sqrt \frac{N_c}{\pi} }  \int_0^1 {\mathrm d} x  \left[ 1 + { m_{s}^{2}-\nabla_{r}^{2} \over x (1-x) M^{2}_{V} } \right] \left. \Psi^\phi_L(x, \zeta) \right|_{r=0}
 \label{fvL}
 \eea
 and
 \begin{equation}
 	f_{\phi}^{\perp}(\mu) =\sqrt{\frac{N_c}{2\pi}} m_s \int_0^1 {\mathrm d} x \; \int {\mathrm d} r \; \mu J_1(\mu r)  \frac{\Psi^\phi_T(x, \zeta)}{x(1-x)}
 	\label{fvT}
 \end{equation}
 respectively.  The vector coupling is also referred to as the decay constant as it is related to the measured electronic decay width $\Gamma_{V \rightarrow e^+ e^-}$ of the vector meson:
 \be \Gamma_{\phi \rightarrow e^+ e^-}={ 4 \pi  \alpha_{em}^2  \over 27 M_\phi }f_\phi^2  \,.
 \label{width-fv}
 \ee
 Using the experimental measurement $\Gamma_{\phi \rightarrow e^+ e^-}=1.263\pm 0.15$ KeV \cite{Tanabashi:2018oca},  and the running of the fine structure constant  below 1 GeV \cite{KLOE-2:2016mgi}, we obtain $f_\phi=225\pm 2$ MeV. In Table \ref{tab:fv-phi}, we compare our predictions for the decay constants with those obtained from lattice QCD and other hadronic models as well as the available experimental data.  We note that  we underestimate the electronic decay width due to the fact that there are likely perturbative corrections that must be taken into account when predicting the electronic decay width.

 \begin{table}
 	\centering
 	\begin{tabular}{c|c|c|c|c}
 		\hline\hline
 		Reference & Approach & $f_{\phi}$ [MeV] & $f_{\phi}^{\perp}$ [MeV] & $f_{\phi}^{\perp}/f_{\phi}$\\ \hline
 		This paper & LF holography & $190\pm 20$  & $150^{+10}_{-20}$ & $0.79\pm 0.13$\\ \hline
 		PDG \cite{Tanabashi:2018oca}  & Exp. data  & $225\pm 2$  & $$ & $$\\ \hline
 		Ref. \cite{Ball:1998kk} & Sum Rules & $254 \pm 3$ & $204 \pm 14$ &   \\ \hline
 		Ref. \cite{Becirevic:2003pn} & Lattice (continuum) &  &  & $0.76 \pm 0.01$ \\ \hline
 		Ref. \cite{Braun:2003jg} & Lattice (finite) &  &  & $0.780 \pm 0.008$ \\ \hline
 		Ref. \cite{Jansen:2009hr} & Lattice (unquenched) &  & $$ & $$ \\ \hline
 		
 		Ref. \cite{Gao:2014bca} & Dyson-Schwinger & $190$ & $150$ & $0.79$ \\ \hline
 	\end{tabular}
 	\caption{Our predictions for the longitudinal and transverse decay constants and their ratio for the $\phi$ meson using $\kappa =0.54 \pm 0.02$ GeV and $m_{s}=0.40\pm 0.15$ GeV.}
 	\label{tab:fv-phi}
 \end{table}

\section{Distribution Amplitudes for the $\phi$}
We now proceed to predict the twist-$2$ DAs $\phi_{K^*}^{\parallel ,\perp} (x,\; \mu )$ given by\cite{Ahmady:2013cva}:
\begin{equation}
	f_{\phi} \phi_\parallel(x,\mu) =\sqrt{\frac{N_c}{\pi}} \int {\mathrm d}
	b \mu
	J_1(\mu b) \left[1 + \frac{ m_{s}^2 -\nabla_b^2}{M_{\phi}^2 x(1-x)}\right] \frac{\Psi^\phi_L(x, \zeta)}{x(1-x)} \;,
	\label{phiparallel-phiL}
\end{equation}
and 
\begin{equation}
	f_{\phi}^{\perp}(\mu) \phi_\perp(x,\mu) =m_s\sqrt{\frac{N_c }{2 \pi}} \int {\mathrm d}
	b \mu
	J_1(\mu b) \frac{\Psi^\phi_T(x,\zeta)}{x(1-x)} \; ,
	\label{phiperp-phiT}
\end{equation}

Note that the decay constants are defined in Eqns \ref{fvL} and \ref{fvT} so that the twist-2 DAs satisfy the  normalization condition, i.e.
\be
\int \phi_{\parallel ,\perp}(x,\; \mu ) dx=1\; .
\ee

\begin{figure}[]
	\begin{subfigure}{}
	\centering
	\includegraphics[trim=2cm 15cm 3.5cm 2cm, clip, width=0.5\textwidth]{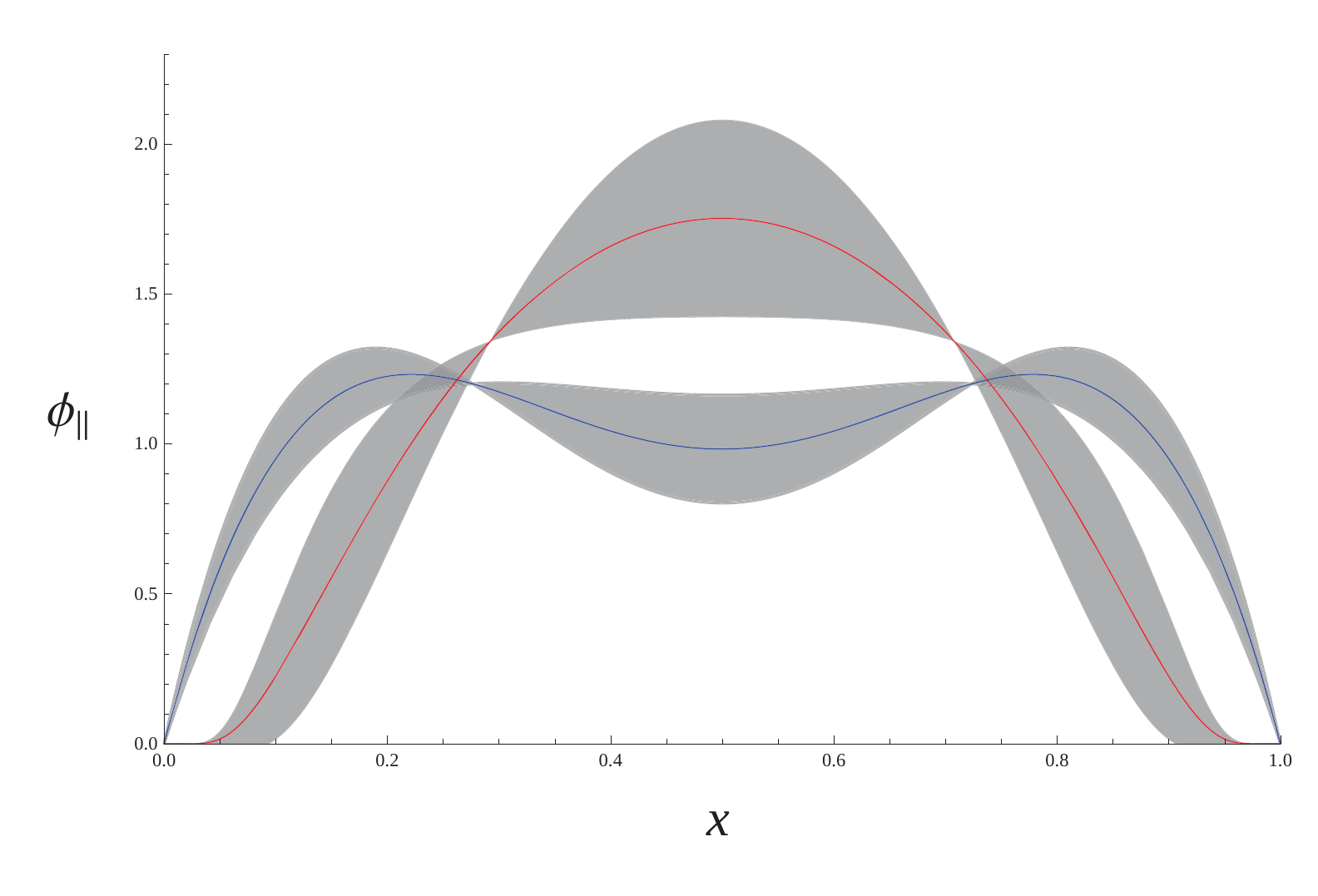}
	\label{adsdapara}
\end{subfigure}
\hspace{.1cm}
\begin{subfigure}{}
	\centering
	\includegraphics[trim=2cm 15cm 3.5cm 2cm, clip, width=0.5\textwidth]{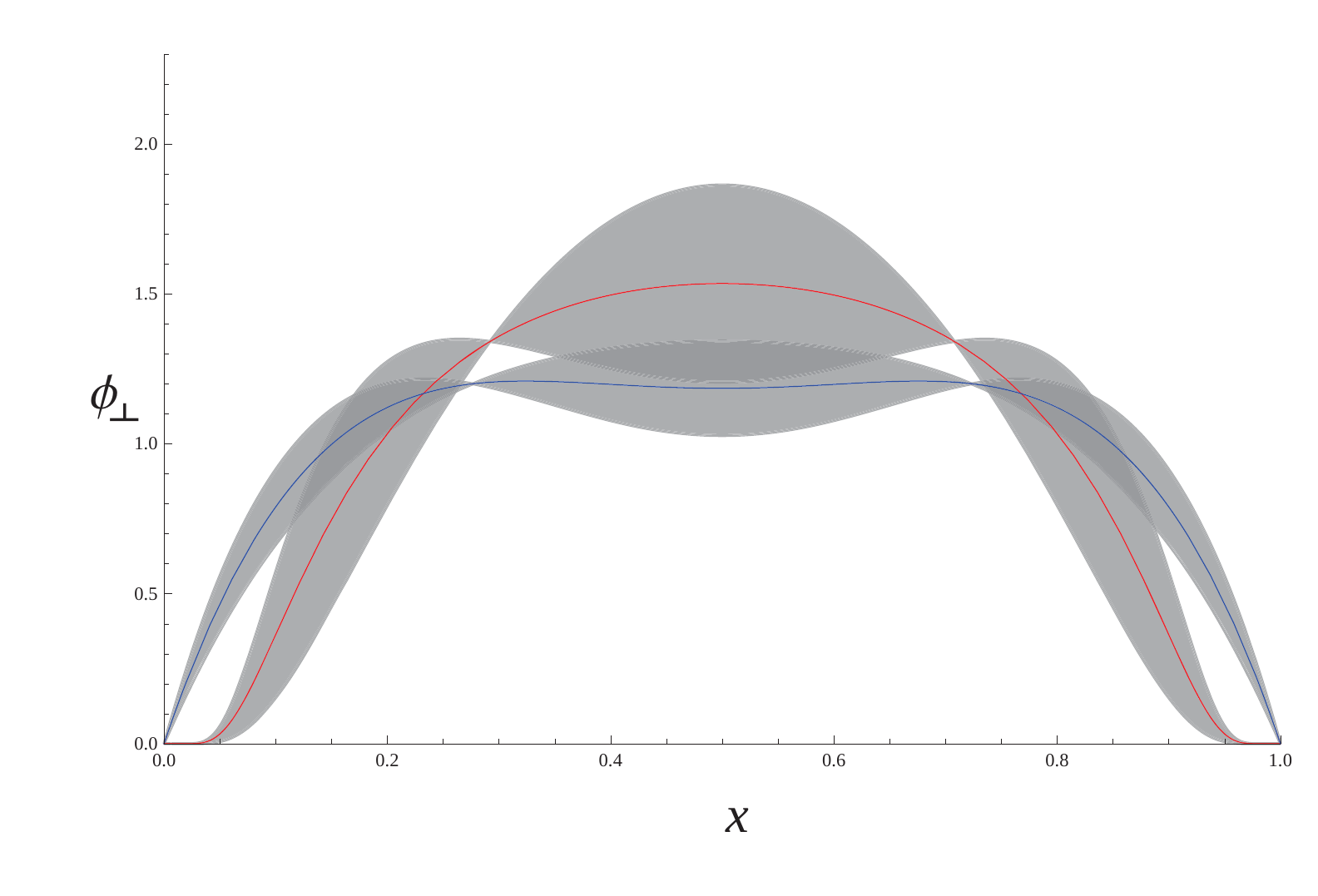}
	\label{srdapara}
\end{subfigure}\\

	\caption{Twist-2 DAs predicted by hQCD (red curve) and SR (blue curve).  The uncertainty band is due to the variation of the quark mass and the fundamental scale for holographic QCD and the error bar on Gegenbauer coefficients for SR.}
	\label{das}
\end{figure}

We can now compare the holographic DAs with those obtained using QCDSR. QCDSR predict the moments of the DAs: 
\begin{equation}
\langle \xi_{\parallel, \perp}^n \rangle_\mu = \int {\mathrm d} x \; \xi^n \phi_{\parallel,\perp} (x, \mu)  
\end{equation}
and that only the first two moments are available in the standard SR approach \cite{Ball:2007zt}. The twist-$2$ DA are then reconstructed as a Gegenbauer expansion
\begin{equation}
\phi_{\parallel,\perp}(x, \mu) = 6 x \bar x \left\{ 1 + \sum_{j=1}^{2}
a_j^{\parallel,\perp} (\mu) C_j^{3/2}(2x-1)\right\} \;. 
\label{phiperp-SR}
\end{equation}
where $C_j^{3/2}$ are the Gegenbauer polynomials and the coeffecients $a_j^{\parallel,\perp}(\mu)$ are related to the moments $\langle \xi_{\parallel,\perp}^n \rangle_\mu$ \cite{Choi:2007yu}. These moments and coefficients are determined at a starting scale $\mu=1$ GeV and can then  be evolved perturbatively to higher scales \cite{Ball:2007zt}.  For mesons with definite G parity (equal mass quark anti-quark), i.e. $\phi$ in our case, $a_1^{\parallel \perp}=0$ \cite{Ball_2007}.  For the other two coefficients, we adopt $a_2^\parallel =0.23\pm 0.08$ and $a_2^\perp =0.14\pm 0.07$ from references \cite{Straub:2015ica,Ball_2007}.  In Fig. \ref{das}, we show the predictions of holographic QCD for twist-2 DAs compared with  those obtained from QCDSR.

Figure \ref{das} shows twist-2 DAs $\phi_{\parallel ,\perp}(x,\; \mu=1\; {\rm GeV})$ for the $\phi$ vector meson obtained using Eqs. \ref{phiparallel-phiL} and \ref{phiperp-phiT} as compared to SR predictions as given by Eq. \ref{phiperp-SR}.  The uncertainty band for holographic DAs are due to the uncertainties in quark mass and the fundamental AdS/QCD scale: $m_s=0.40\pm 0.15$ GeV and $\kappa =0.54\pm 0.02$ GeV. The error band in SR DAs are the result of the uncertainties in the Gegenbauer coefficients as given above.  

The twist-3 DAs can also be obtained from the light-front wavefunction through the following expressions \cite{Ahmady:2013cva}

\begin{equation}
g_\perp^{(v)}(x)=\frac{N_c}{2\pi f_\phi m_\phi}\int {\mathrm d}
b \mu
J_1(\mu b) [m_s^2-(x^2+(1-x)^2\nabla_b^2)]\frac{\Psi^\phi_T(x,\zeta)}{x^2(1-x)^2} \; ,
\label{gv}
\end{equation}
\begin{equation}
\frac{{\mathrm d}g_\perp^{(a)}(x)}{{\mathrm d}x}=\frac{\sqrt{2}N_c}{\pi f_\phi m_\phi}\int {\mathrm d}
b \mu
J_1(\mu b) [(1-2x)m_s^2-\nabla_b^2)]\frac{\Psi^\phi_T(x,\zeta)}{x^2(1-x)^2} \; ,
\label{ga}
\end{equation}
\begin{figure}[]
	\begin{subfigure}{}
		\centering
		\includegraphics[trim=2cm 15cm 3.5cm 2cm, clip, width=0.5\textwidth]{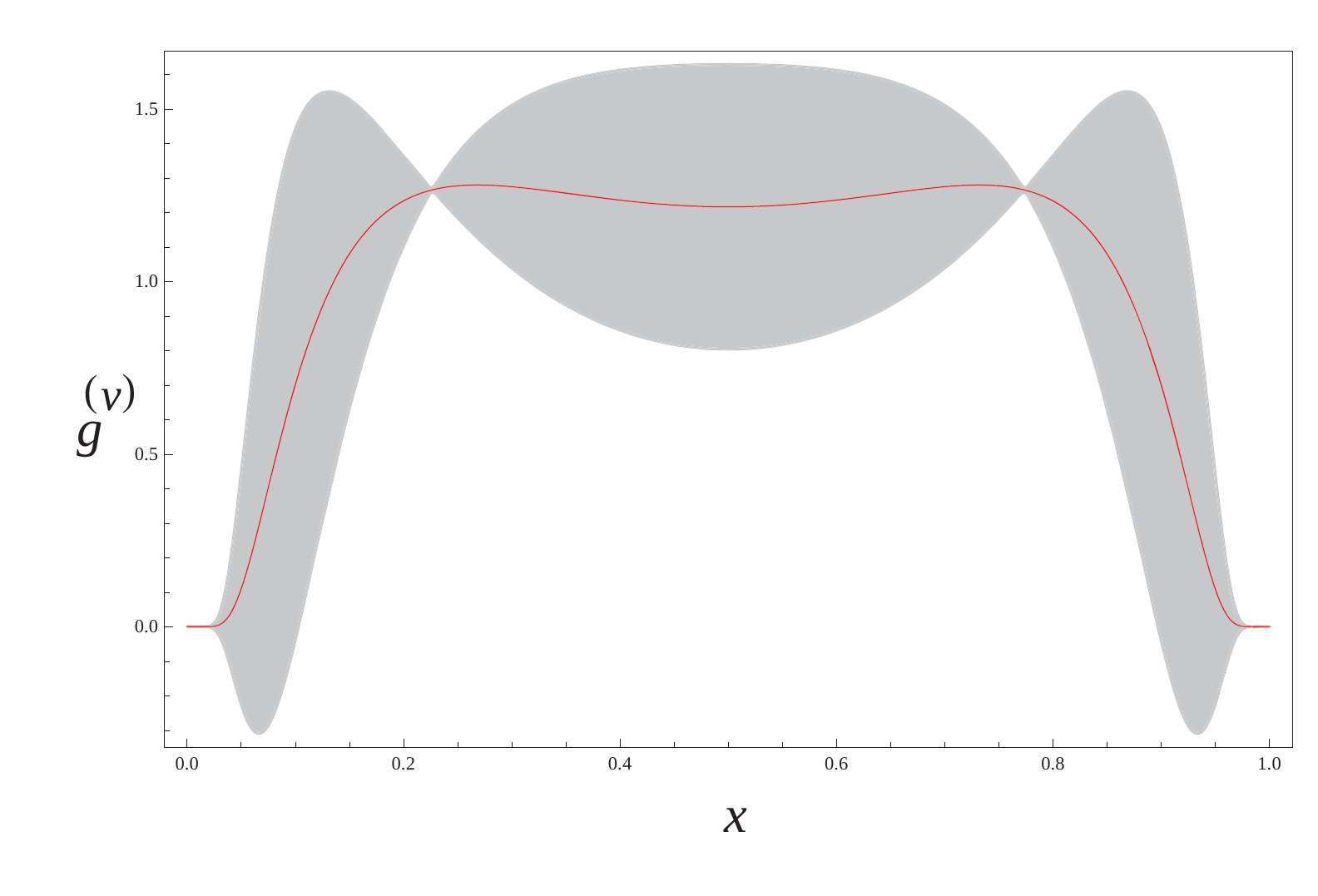}
		\label{adsdapara}
	\end{subfigure}
	\hspace{.1cm}
	\begin{subfigure}{}
		\centering
		\includegraphics[trim=2cm 15cm 3.5cm 2cm, clip, width=0.5\textwidth]{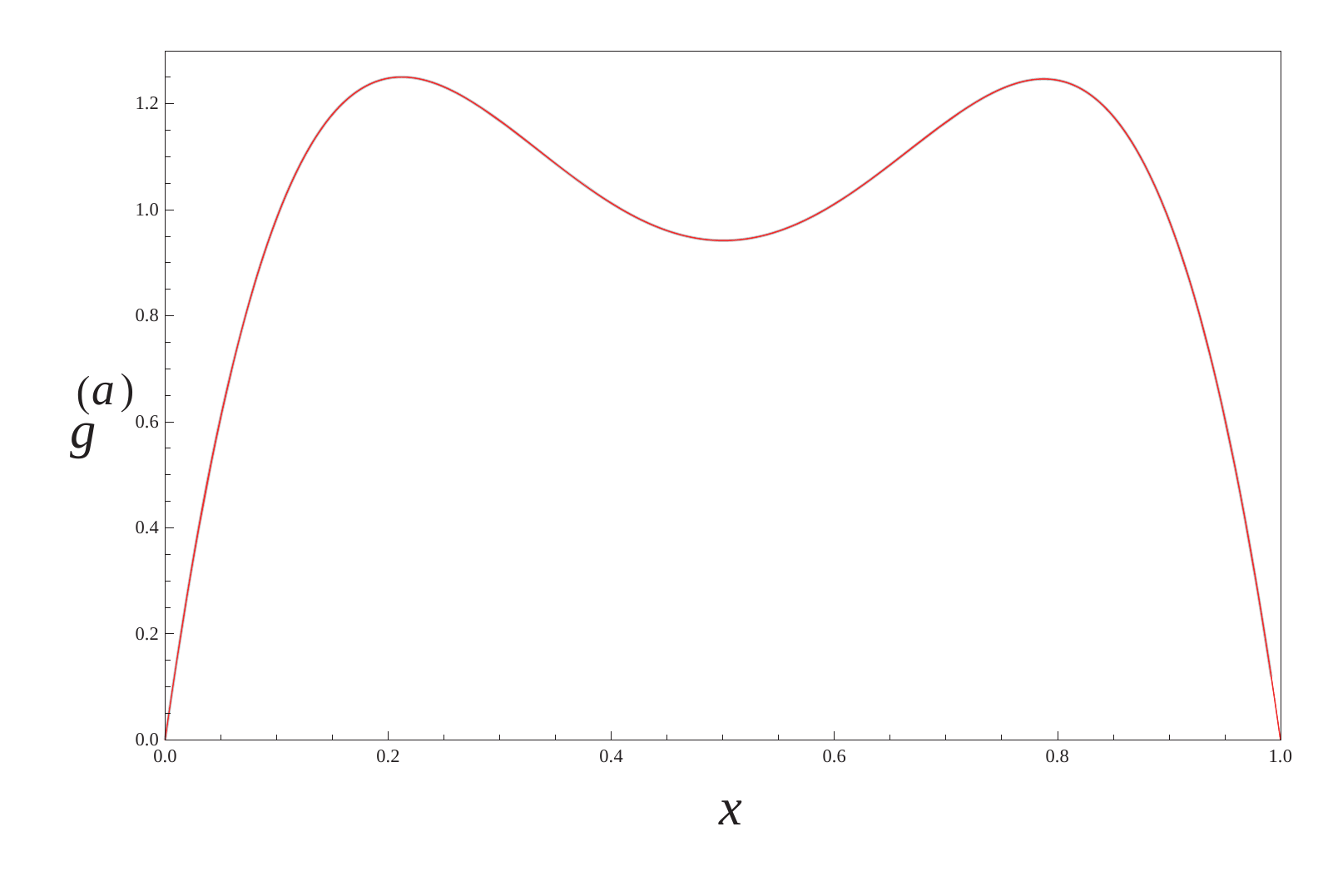}
		\label{srdapara}
	\end{subfigure}\\
	
	\caption{Twist-3 DAs $g^{(v)}$ and $g^{(a)}$ predicted by hQCD.  The uncertainty band is due to the variation of the quark mass and the fundamental scale for holographic QCD.}
	\label{T3DAs}
\end{figure}
Figure \ref{T3DAs} shows the twist-3 DAs for $\phi$ meson which are obtained from Eqns \ref{gv} and \ref{ga}.  As is evident from the figures, the uncertainty due to $m_s$ and $\kappa$, contrary to the case for $g^{(v)}$, is negligible for $g^{(a)}$.  In the next section, we use the decay constants and DAs up to twist-3 to compute the transition form factors (TFFs) $B_s\to\phi$.
\section{$B_s\rar \phi$ transition form factors}
  The seven $B_s\to\phi$ TTFs are defined as\cite{Horgan:2013hoa}
  \begin{eqnarray}
  \langle \phi (k,\varepsilon)|\bar{s} \gamma^\mu(1-\gamma^5 )b | B_s(p) \rangle &=& \frac{2i V(q^2)}{m_{B_s} + m_{\phi}} \epsilon^{\mu \nu \rho \sigma} \varepsilon^*_{\nu} k_{\rho} p_{\sigma} -2m_{\phi} A_0(q^2) \frac{\varepsilon^* \cdot q}{q^2} q^{\mu}  \nonumber \\
  &-& (m_{B_s} + m_{\phi}) A_1(q^2) \left(\varepsilon^{\mu *}- \frac{\varepsilon^* \cdot q q^{\mu}}{q^2} \right) \nonumber \\
  &+& A_2(q^2) \frac{\varepsilon^* \cdot q}{m_{B_s} + m_{\phi}}  \left[ (p+k)^{\mu} - \frac{m_{B_s}^2 - m_{\phi}^2}{q^2} q^{\mu} \right] ,
  \label{vaff}
  \end{eqnarray}
  \begin{eqnarray}
  q_{\nu} \langle \phi (k,\varepsilon)|\bar{s} \sigma^{\mu \nu} (1+\gamma^5 )b | B_s(p) \rangle &=& 2 T_1(q^2) \epsilon^{\mu \nu \rho \sigma} \varepsilon^*_{\nu} p_{\rho} k_{\sigma} \nonumber \\
  &+& i T_2(q^2)[(\varepsilon^* \cdot q)(p+k)_{\mu}-\varepsilon_{\mu}^*(m_{B_s}^2-m_{\phi}^2)] \nonumber \\
  &+& iT_3(q^2) (\varepsilon^* \cdot q) \left[ \frac{q^2}{m_{B_s}^2-m_{\phi}^2} (p+k)_{\mu} -q_{\mu}  \right] ,
  \label{tff}
  \end{eqnarray}
  where $q=p-k$ is the 4-momentum transfer and $\epsilon$ is the polarization 4-vector of the $\phi$.  At low-to-intermediate values of $q^2$, these TFFs can be computed using QCD light-cone sum rules (LCSR) \cite{Ali:1993vd,Aliev:1996hb,Ball:2004rg}.  Here we use the LCSR expressions from Ref.\cite{Ahmady:2014sva} which are modified for the $B_s\to\phi$ decay channel. Table \ref{inputs} shows the numerical values of the input parameters used in our predictions of the TFFs and the decay rate.
   \begin{table}[h]
   	\begin{tabular}{|c|c|c|c|}
   		\hline $m_{B_s}$& $5.367\; \mathrm{GeV}$ & $M_{B}$ & $8\;\mathrm{GeV}^2$   \\ 
   		\hline $\alpha^{-1}$ & $127$ & $s_0$ & $36\;\mathrm{GeV}^2$   \\ 
   		\hline	$|V_{tb}V_{ts}^{*}|$ & $0.0407$ & $f_{B_s}$ & $0.224\;\mathrm{GeV}$   \\ 
   		\hline	$\tau_{B_s}$ & $1.512\; \mathrm{ps}$ & $m_b$ & $4.8\;\mathrm{ GeV}$   \\  
   		\hline 
   	\end{tabular}
   	\caption{Numerical values of the input parameters.}
   	\label{inputs}
   \end{table}
  The form factors, computed via LCSR, are valid at low to intermediate $q^2$.  The extrapolation to high $q^2$ is performed via a two-parameter fit of the following form
\begin{equation}
F(q^2)=\frac{F(0)}{1-a(q^2/m_{B_s}^2)+b(q^4/m_{B_s}^4)}
\end{equation}
to the LCSR predictions as well as form factor values obtained by lattice QCD which are available at high $q^2$.  The results for the above fit are given in Table \ref{table:formfitslattice}.  We note that hQCD predicts lower values of the form factors at $q^2=0$ compared to those obtained from SR.

\begin{table}[h]
	\begin{tabular}{|c|c|c|c|c|c|c|c|}
		\hline  & V & $A_0$ & $ A_1$ & $A_2$ & $T_1$ & $T_2$ & $T_3$ \\
		\hline   F(0) (hQCD)&$0.26\pm 0.2$  &$0.15\pm 0.03$  &$0.19 \pm 0.02$  &$0.21\pm 0.01$  &$0.21\pm 0.03$ &$0.22\pm 0.01$ &$0.16\pm 0.01$\\ 
		\hline  F(0) (SR) &$0.33\pm 0.01$  &$0.27\pm 0.01$  &$0.26\pm 0.01$  &$0.26\pm 0.01$  &$0.27\pm 0.01$ &$0.28\pm 0.01$ &$0.18\pm 0.01$\\ 
		\hline  a (hQCD) & $1.66\pm 0.06$ &$2.21^{+0.18}_{-0.15}$ & $0.85^{+0.08}_{-0.07}$ &$1.41\pm 0.22$ & $1.76^{+0.20}_{-0.16}$ & $0.51\pm 0.01$& $0.55^{+0.32}_{-0.42}$\\ 
		\hline  a (SR) &  $1.50\pm 0.01$ & $1.26\pm 0.03$ & $0.24\pm 0.02$& $1.35^{+0.31}_{-0.12}$ & $1.51\pm 0.02$ & $0.046\pm 0.01$& $0.73\pm 0.19$\\ 
		\hline   b (hQCD) &$0.60^{+0.08}_{-0.07}$  &$1.26^{+0.24}_{-0.19}$  &$-0.23^{+0.07}_{-0.06}$  &$0.51^{+0.66}_{-0.13}$  &$0.74^{+0.25}_{-0.20}$  &$-0.63\pm 0.05$ &$-0.98^{+0.44}_{-0.55}$\\
		\hline  b (SR)  &$0.45\pm 0.01$  &$-0.013\pm 0.048$  &$-0.86\pm 0.03$  &$0.62^{+0.85}_{-0.39}$  &$0.48\pm 0.03$ &$-1.08\pm 0.03$ &$-0.65^{+0.23}_{-0.19}$\\
		\hline
	\end{tabular}
	\caption{hQCD+ lattice predictions for the form factors. Lattice data are taken from \cite{Horgan:2013hoa}.  The error bars for the holographic form factors are due to the variation in $m_s$ and $\kappa$, as explained in the text.}
	\label{table:formfitslattice}
\end{table}  

\begin{figure}[htbp]
	\begin{subfigure}{}
	\centering
	\includegraphics[width=0.5\textwidth]{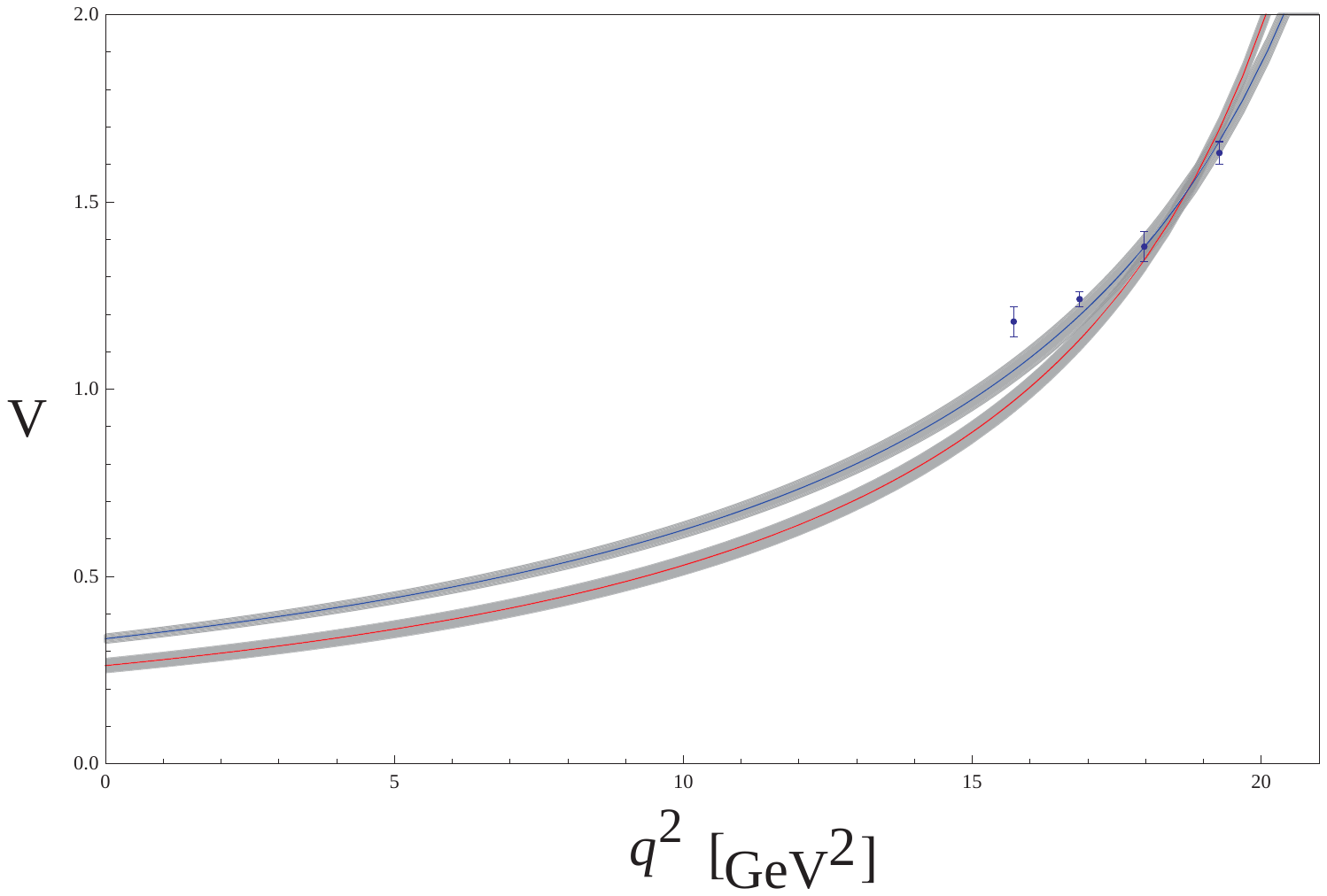}
	\label{fig:plot_V}
\end{subfigure}
\hspace{-2.1cm}
\begin{subfigure}{}
	\centering
	\includegraphics[width=0.5\textwidth]{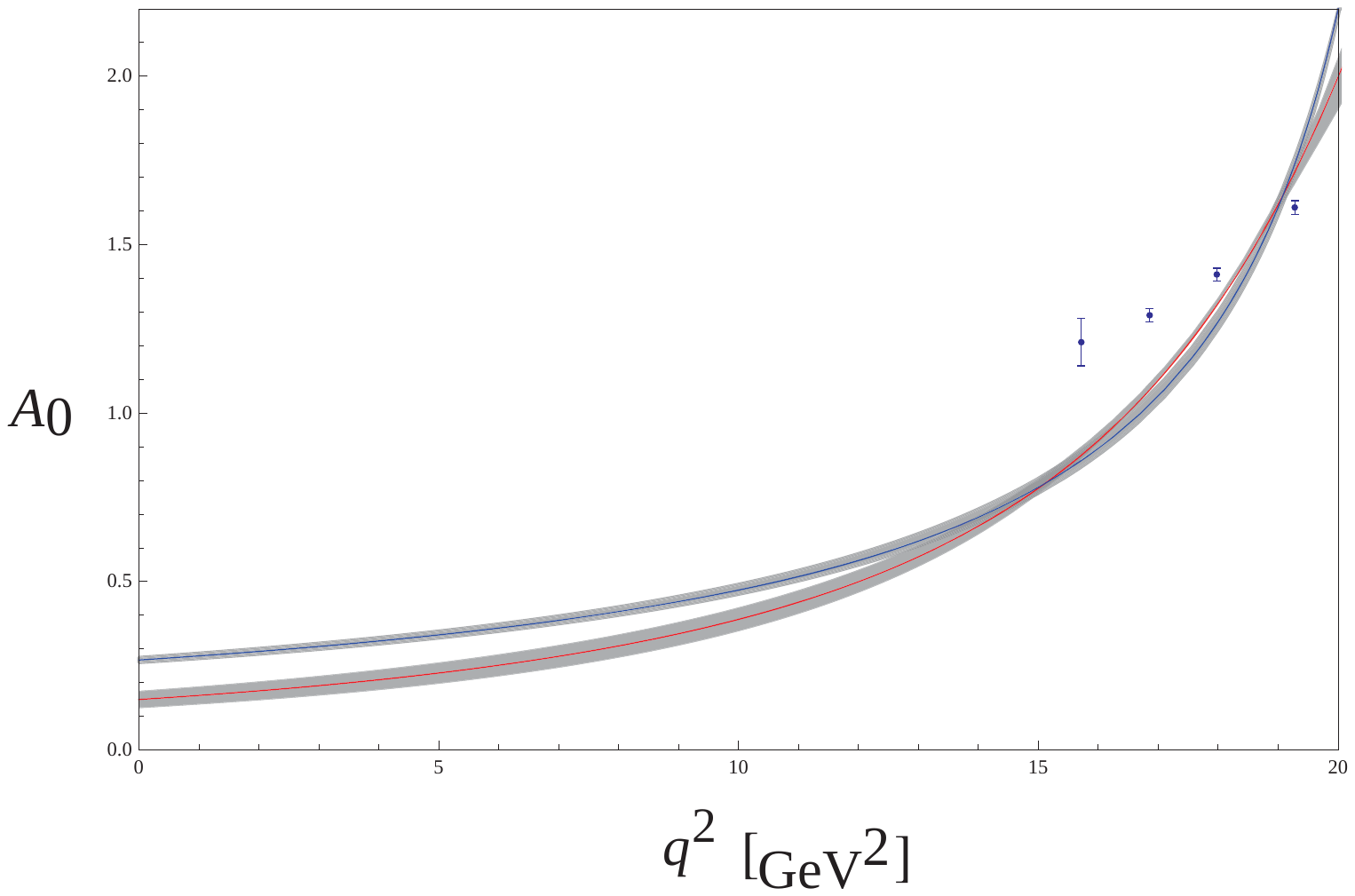}
	\label{fig:plot_A0}
\end{subfigure}
	\vspace{-6.1cm}
	\begin{subfigure}{}
		\centering
			\includegraphics[width=0.5\textwidth]{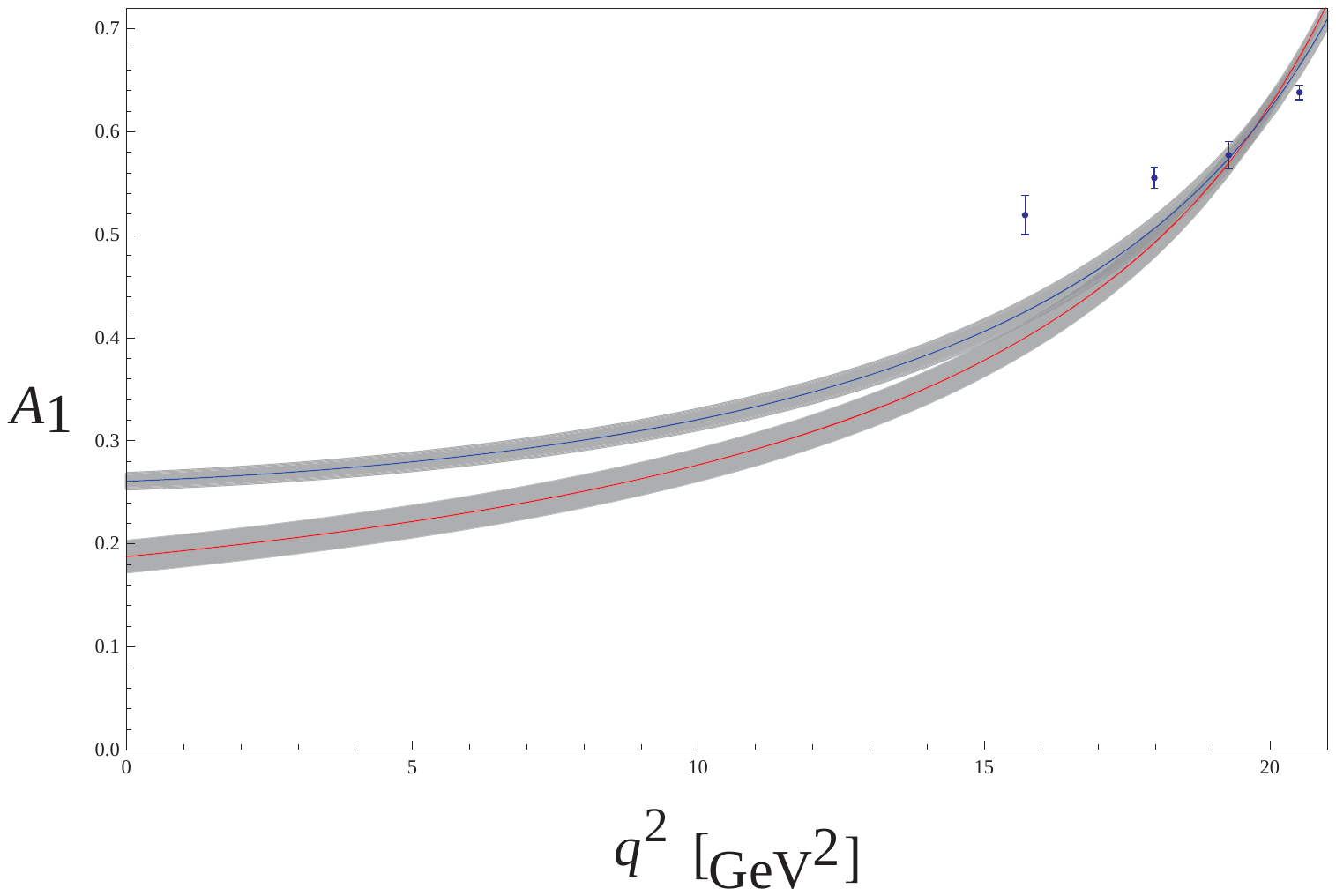}
		\label{fig:plot_A1}
	\end{subfigure}
		\hspace{-2.1cm}
		\begin{subfigure}{}
			\centering
			\includegraphics[width=0.5\textwidth]{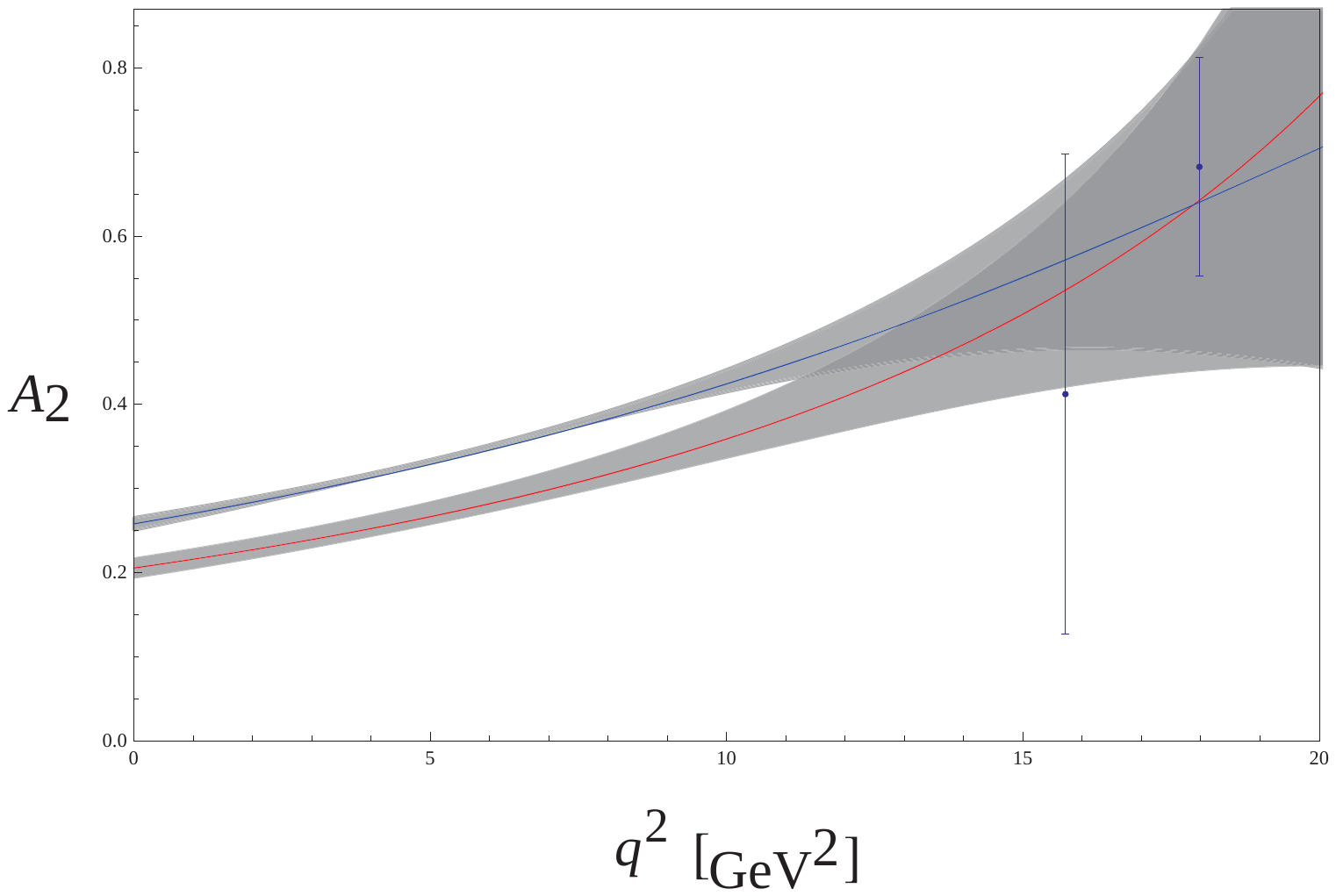}
			\label{fig:plot_A2}
		\end{subfigure}
		\vspace{-6.1cm}
		\begin{subfigure}{}
			\centering
			\includegraphics[width=0.5\textwidth]{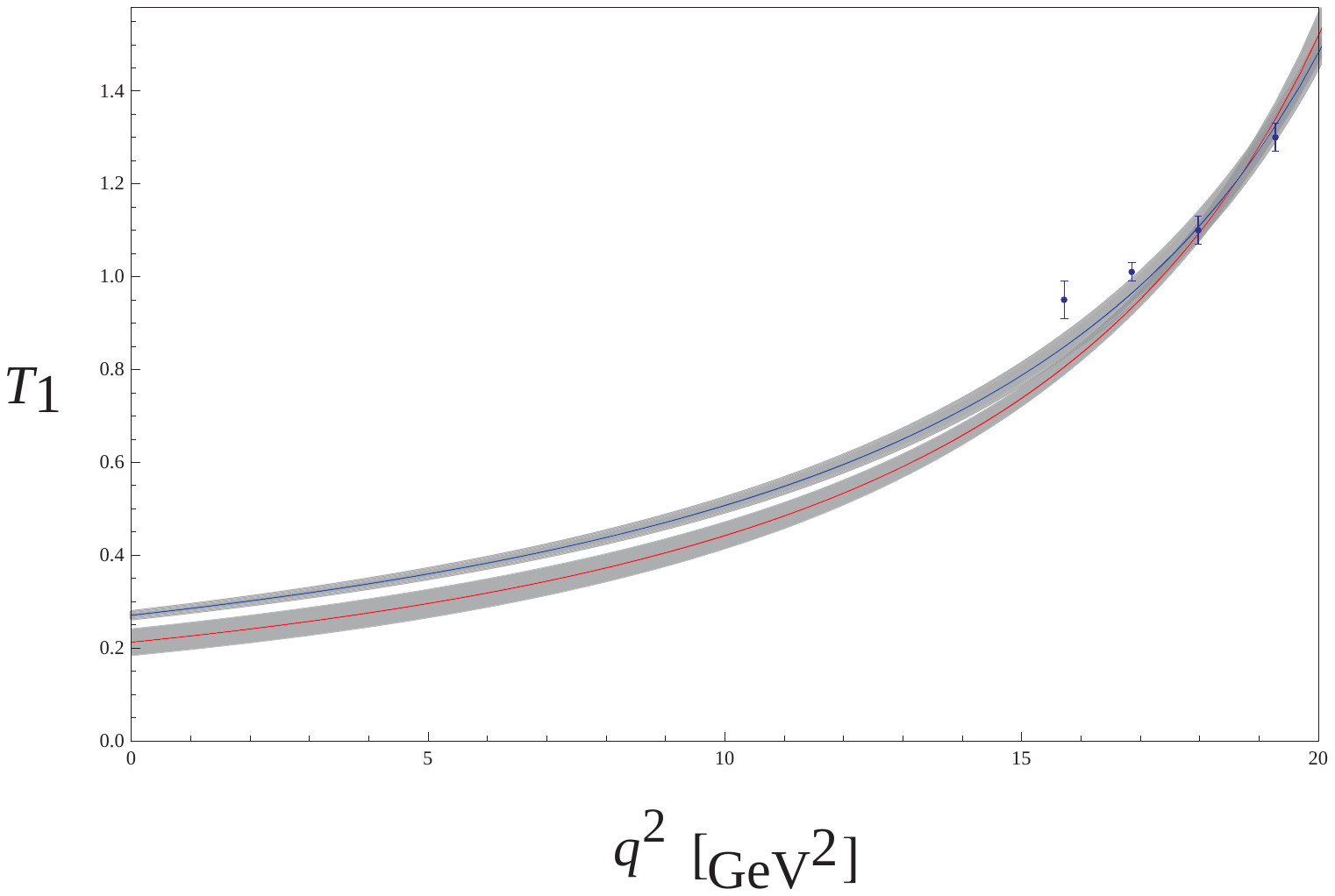}
			\label{fig:plot_T1}
			\end{subfigure}
			\hspace{-2.1cm}	
			\begin{subfigure}{}
				\centering
				\includegraphics[width=0.5\textwidth]{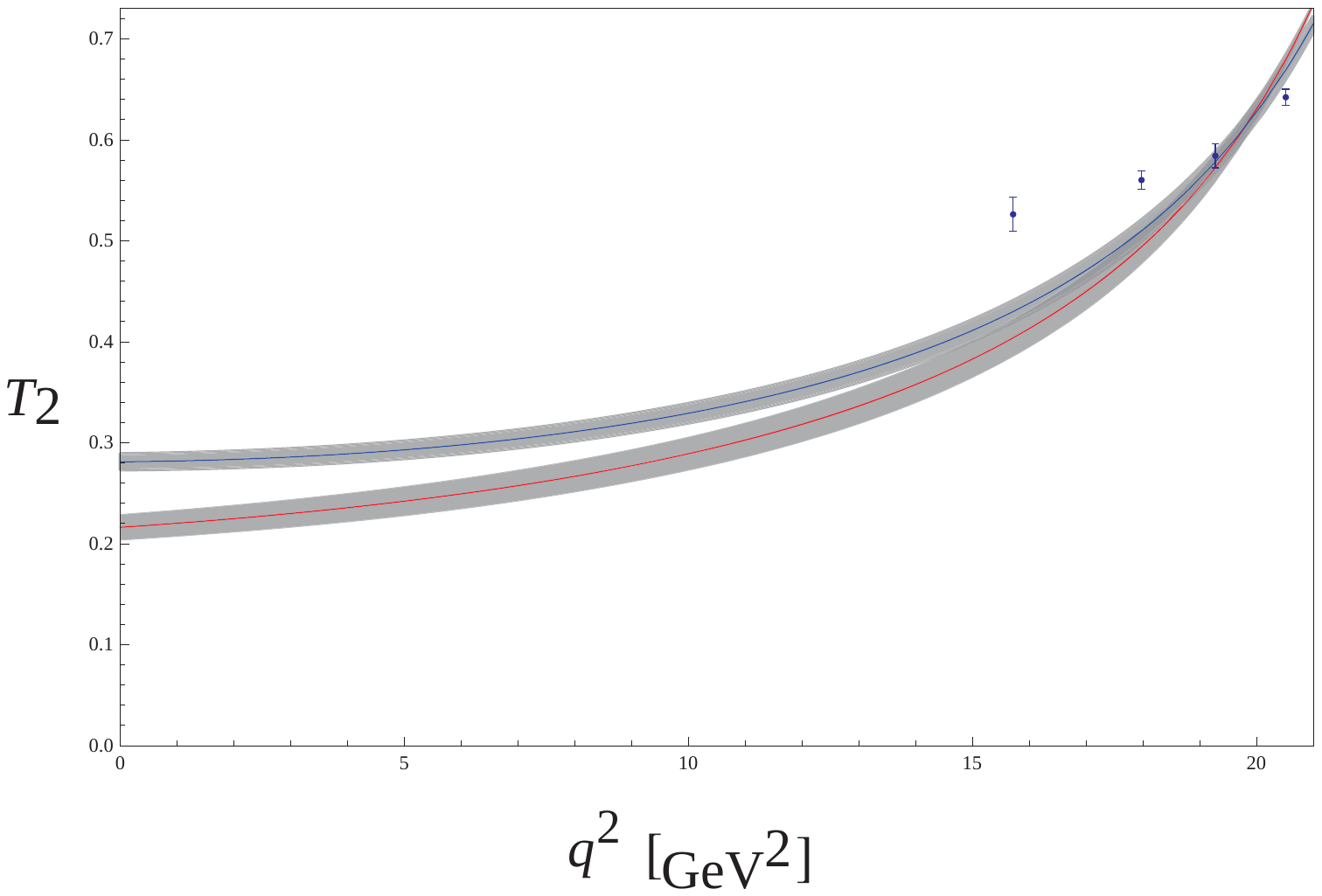}
				\label{fig:plot_T2}				
\end{subfigure}
\vspace{-6.1cm}
\begin{subfigure}{}
	\centering
	\includegraphics[width=0.5\textwidth]{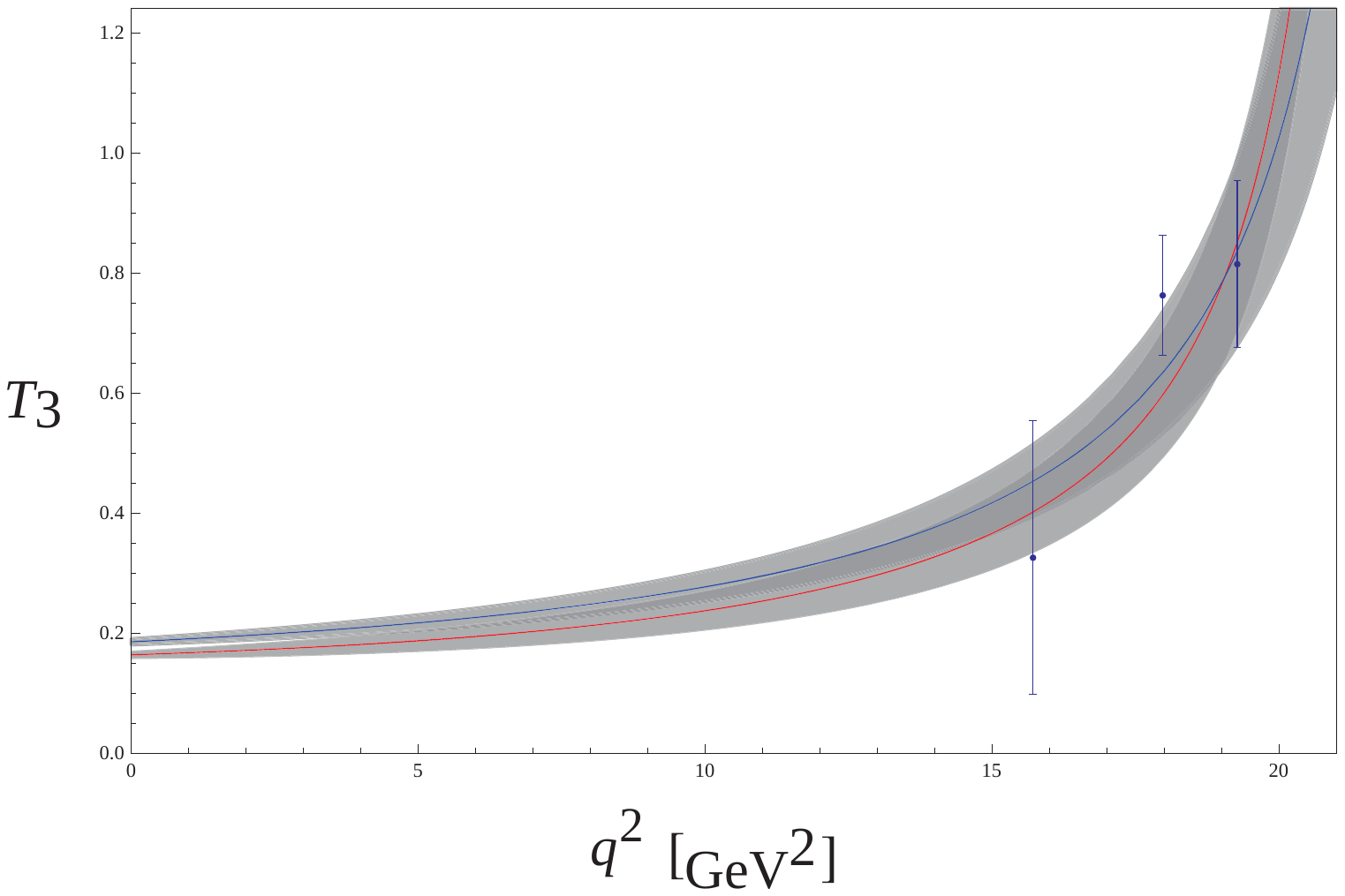}
	\label{fig:plot_T3}
\end{subfigure}
\vspace{-5.1cm}
	\caption{hQCD (red curves) and QCDSR (blue curves) predictions for the $B_s\to\phi$ TFFs. The two-parameter fits with the available lattice data are used for the plots.  The shaded bands represent the uncertainty in the predicted form factors due to uncertainty bands in DAs and variation in quark masses.  Lattice data points (shown on plots) are taken from \cite{Horgan:2013hoa}. }
	\label{formfactors}
\end{figure}

Figure \ref{formfactors} shows the comparison between hQCD and SR predictions including the lattice data points at high $q^2$ for the form factors.  The shaded bands in these figures represent the uncertainty due to the error band in the DAs.  Note that there are additional uncertainties in the form factors inherent in the LCSR method (uncertainty in the Borel parameter, continuum threshold and other input parameters).  Since our goal in this paper is to discriminate between the hQCD and SR models and that the inherent LCSR uncertainties are the same in both models, we do not include them here.

\section{Differential decay rate}
\begin{figure}
	\includegraphics[trim=1cm 15cm 3.5cm 2cm, clip, width=0.7\textwidth]{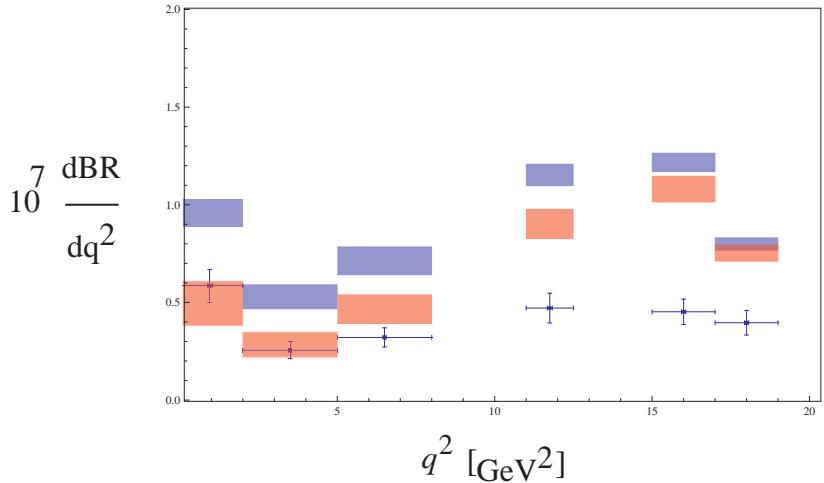}
	\caption{The differential branching ratio for \processphi as predicted by hQCD (red rectangles) and SR (blue rectangles).  The uncertainty widths are due to the form factors.  The experimental data points are measured by LHCb \cite{Aaij:2015esa}.}
	\label{DBRatio}
\end{figure}

The differential branching ratio for $B_s\to\phi\mu^+\mu^-$ is given by the following expression \cite{Aliev:1996hb}:
\begin{eqnarray}\label{gammaK}\nonumber
	&&\frac{d\mathcal{B}}{dq^{2}}=\tau_{B_s}\frac{G_{F}^{2}\alpha^{2}}{2^{11}\pi^{5}}\frac{|{V_{tb}V_{ts}^{*}|^{2}\sqrt{\lambda}v}}{3m_{B_s}}((2m_{\mu}^{2}+m_{B_s}^{2}s)[16(|A|^{2}+|C|^{2})m_{B_s}^{4}\lambda+2(|B_{1}|^{2}+|D_{1}|^{2})\\ \nonumber
	&&\times\frac{\lambda+12rs}{rs}+2(|B_{2}|^{2}+|D_{2}|^{2})\frac{m_{B_s}^{4}\lambda^{2}}{rs}-4[\Re \mbox{e}(B_{1}B_{2}^{*})+\Re\mbox{e}(D_{1}D_{2}^{*})]\frac{m_{B_s}^{2}\lambda}{rs}(1-r-s)]\\ \nonumber
	&&+6m_{\mu}^{2}[-16|C|^{2}m_{B_s}^{4}\lambda+4\Re \mbox{e}(D_{1}D_{3}^{*})\frac{m_{B_s}^2\lambda}{r}-4\Re \mbox{e}(D_{2}D_{3}^{*})\frac{m_{B_s}^{4}(1-r)\lambda}{r}+2|D_{3}|^{2}\frac{m_{B_s}^{4}s\lambda}{r}\\ \nonumber
	&&-4\Re \mbox{e}(D_{1}D_{2}^{*})\frac{m_{B_s}^{2}\lambda}{r}-24|D_{1}|^{2}+2|D_{2}|^{2}\frac{m_{B_s}^{4}\lambda}{r}(2+2r-s)])
\end{eqnarray}
where $\lambda=1+r^{2}+s^{2}-2r-2s-2rs$, with $r=m_{\phi}^{2}/m_{B_s}^{2}$ and $s=q^{2}/m_{B_s}^{2}$. The final muon has  mass $m_{\mu}$ and velocity $v=\sqrt{1-4 m_{\mu}^{2}/q^{2}}$. The differential branching fraction depends on the following combinations of form factors: 

\begin{eqnarray}\label{A}
	A=C_{9}^{eff}\left(\frac{V}{m_{B_s}+m_{\phi}}\right)+4C_{7}\frac{m_{b}}{q^{2}}T_{1} \;,
\end{eqnarray}

\begin{eqnarray}\label{B1}
	B_{1}=C_{9}^{eff}(m_{B_s}+m_{\phi})A_{1}+4C_{7}\frac{m_{b}}{q^{2}}(m_{B_s}^{2}-m_{\phi}^{2})T_{2} \;,
\end{eqnarray}

\begin{eqnarray}\label{B2}
	B_{2}=C_{9}^{eff}\left(\frac{A_{2}}{m_{B_s}+m_{\phi}}\right) + 4C_{7}\frac{m_{b}}{q^{2}}\left(T_{2}+\frac{q^{2}}{m_{B_s}^{2}-m_{\phi}^{2}}T_{3}\right) \;,
\end{eqnarray}

\begin{eqnarray}\label{C}
	C=C_{10}\left(\frac{V}{m_{B_s}+m_{\phi}} \right)\;,
\end{eqnarray}

\begin{eqnarray}\label{D1}
	D_{1}=C_{10}(m_{B_s}+m_{\phi})A_{1} \;,
\end{eqnarray}

\begin{eqnarray}\label{D2}
	D_{2}=C_{10}\left(\frac{A_{2}}{m_{B_s}+m_{\phi}} \right)\;,
\end{eqnarray}
and
\begin{eqnarray}\label{D3}
	D_{3}=-C_{10}\frac{2m_{\phi}}{q^{2}}\left( A_3-A_{0}\right)\;,
\end{eqnarray}
where 
\begin{eqnarray}\label{A3}
A_3=A_{0}-\frac{q^2}{2m_\phi}\frac{A_2}{m_{B_s}+m_\phi}\;.
\end{eqnarray}
As for the SM Wilson coefficients $C_7$, $C_9^{eff}=C_9 + Y(q^2)$ and $C_{10}$ appearing in the above equations,  are given in  Ref. \cite{Altmannshofer:2008dz} .  Figure \ref{DBRatio} shows the hQCD and QCDSR predictions for \processphi differential branching ratio compared with the available experimental data from LHCb \cite{Aaij:2015esa}.  We observe that at low momentum transfer where the form factors are most sensitive to  DAs through LCSR, hQCD produces better agreement with the data.  The bin-by-bin numerical predictions are given in Table \ref{table:databin}.  

\begin{figure}[]
	\begin{subfigure}{}
		\centering
		\includegraphics[trim=2cm 15cm 3.5cm 2cm, clip, width=0.5\textwidth]{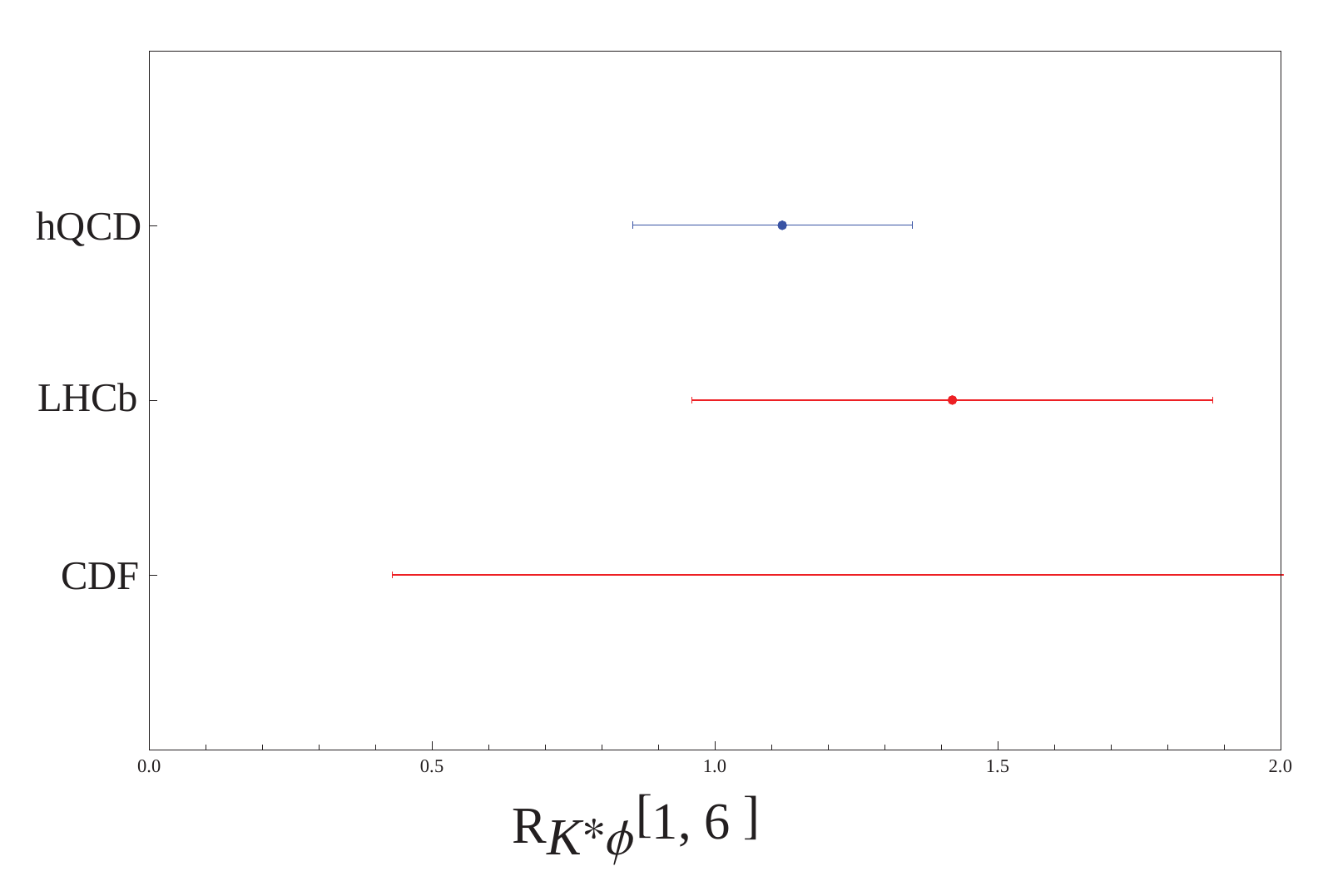}
		\label{LowRatio}
	\end{subfigure}
	\hspace{.1cm}
	\begin{subfigure}{}
		\centering
		\includegraphics[trim=2cm 15cm 3.5cm 2cm, clip, width=0.5\textwidth]{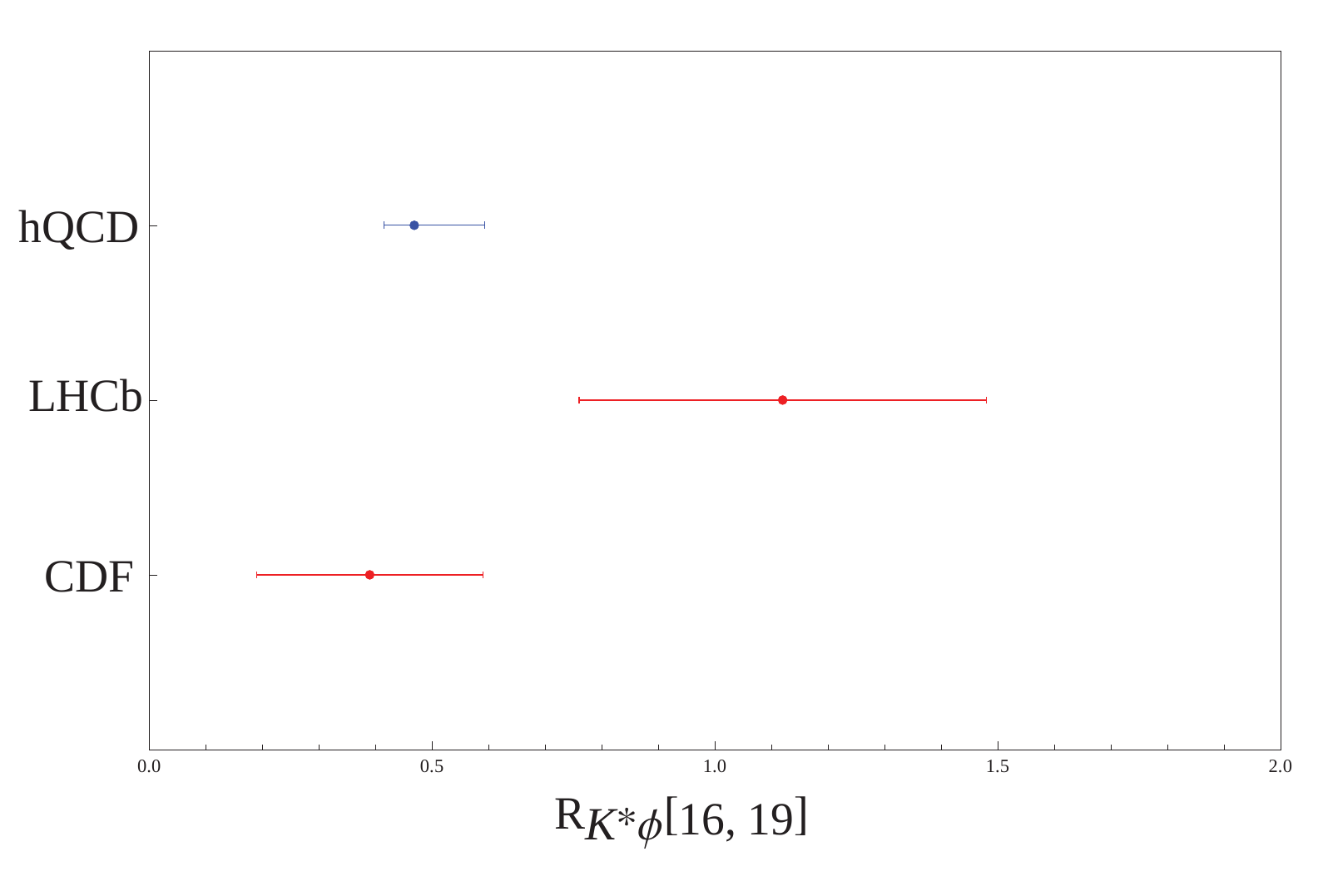}
		\label{HighRatio}
	\end{subfigure}\\
		\caption{hQCD prediction for $R_{K^*\phi}$ at low and hight $q^2$ ranges as compared with the experimental data from LHCb \cite{Aaij:2013iag,Aaij:2014pli} and CDF\cite{Cdf:2012cdf}.}
		\label{Ratiographs}
	\end{figure}

Using our hQCD predictions for \processKstar decay \cite{Ahmady:2015fha}, we can consider the ratio
\begin{equation}
R_{K^*\phi}[q_1,\;q_2]=\frac{dBR(B^0\to K^{*0}\mu^+\mu^-)/dq^2|_{[q_1,\;q_2]}}{dBR(B_s\to \phi\mu^+\mu^-)/dq^2|_{[q_1,\;q_2]}}\;,
\label{ratio}
\end{equation}
by using the differential branching ratios integrated over range $[q_1,\;q_2]$.  Figure \ref{Ratiographs} shows a graphical comparison of our predictions for $R_{K^*\phi}$ to the experimental data of LHCb \cite{Aaij:2013iag,Aaij:2014pli} and CDF \cite{Cdf:2012cdf} at low and high $q^2$ range.  It is encouraging that in low $q^2$, the hQCD prediction agrees, within the error bars, with both LHCb and CDF results.  Again, one should note that LCSR method for the evaluation of the transition form factors are expected to be more reliable at low $q^2$.  On the other hand, at high $q^2$, our prediction only agrees with the CDF datum.
\begin{table}[h]
	\begin{tabular}{|cccc|}
		\hline $q^2$ (GeV) & $10^7 \langle \mathcal{BR}^{hQCD}_{}\rangle$ & $10^7\langle \mathcal{BR}^{QCDSR}\rangle$ & $ 10^7\langle \mathcal{BR}^{experiment}\rangle $ \\ 
		\hline 
		$0.1-2.0$ & $0.49^{+0.12}_{-0.11}$ &$0.96{\pm}0.07$& $0.59\pm 0.07$\\ 
		$2-5$ & $0.29^{+0.05}_{-0.07}$ &$0.54^{+0.05}_{-0.07}$& $0.26\pm 0.04$   \\ 
		$5-8$ & $0.48^{+0.06}_{-0.08}$ &$0.73^{+0.06}_{-0.09}$& $0.32\pm 0.04$ \\ 
		$11.0-12.5$ & $0.91^{+0.07}_{-0.08}$ &$1.15{\pm}0.06$& $0.47\pm {0.07}$   \\ 
		$15-17$ & $1.08^{+0.07}_{-0.06}$ &$1.21\pm 0.05$& $0.45^{+0.06}_{-0.05}$ \\ 
		$17.0-19.0$ & $0.75\pm 0.04$ &$0.80^{+0.04}_{-0.03}$& $0.40_{-0.05}^{+0.06}$  \\ 
		\hline
	\end{tabular}
\caption{Bin-by-bin hQCD and QCDSR predictions for the \processphi branching ratio compared with the experimental data from LHCb \cite{Aaij:2015esa}.}
\label{table:databin}
\end{table}

\section{Conclusion}%
We have calculated the $B_s\to\phi$ transition form factors using $\phi$ meson holographic DAs.  Our prediction for the \processphi differential branching ratio is in better agreement with the latest LHCb data than the prediction generated using QCDSR DAs.  In addition, we found that the hQCD prediction for $R_{K^*\phi}$ is in excellent agreement with both LHCb and CDF data at low $q^2$.  We conclude that it is important to have a better understanding of non-perturbative effects in rare $B_{(s)}$ decays.

\section{Acknowledgments} 
   M. A and R. S are supported by Individual Discovery
   Grants from the Natural Science and Engineering Research
   Council of Canada (NSERC): SAPIN-2017-00033 and
   SAPIN-2017-00031 respectively.  We would like to thank Idriss Amadou Ali for his contribution in the numerical computations of this work.

\bibliographystyle{apsrev}
\bibliography{Bphimumu}

\end{document}